# The Relationship between Information Processing and Meaning Generation among Heterogeneous groups of Investors and the Market Asset Price Movement

Inga Ivanova,[1]

**Abstract**

The paper examines how the market asset's price movement emerges from the information exchange between heterogeneous groups of traders. In this respect financial markets, as complex social systems, are examined from the perspective of communication theory regarding the dynamics of information and meaning. The paper also contributes to the development of the quantity theory of meaning as applied to financial markets. The model is based on the quantitative theory of meaning, which can be considered as a complement to quantitative theory of information. Different groups of traders use different criteria for processing information, so the same information can have different meanings. Meanings shape traders' expectations, which are manifested in the price movements of market assets. These dynamics can be captured using a nonlinear evolutionary equation. Model predictions are tested on a variety of foreign exchange, energy, food and indices markets over different time periods. The results show that the time series predicted by the model are co-integrated with the time series of market asset prices, suggesting that the proposed model can adequately describe some traders' behavior patterns. In a narrow sense the proposed methodology can be used to better understand and predict future price movements of market assets. I a wider sense it helps to better understand the behavior of heterogeneous social systems of different origins since information is a universal concept. From this point of view, the transmission of information in financial markets is not much different from the transmission of information in other social systems. The same conceptual tools can be applied to study other complex social systems with similar topologies.

---

[1] corresponding author; Institute for Statistical Studies and Economics of Knowledge, National Research University Higher School of Economics (HSE University), 20 Myasnitskaya St., Moscow, 101000, Russia; and School of Economics and Management, Far Eastern Federal University, 8, Sukhanova St., Vladivostok, 690990, Russia inga.ivanova@hse.ru;

## I. Introduction

The question of whether financial markets can be predicted is a major concern for traders and policymakers. The same applies to mechanisms to control price movements of market assets. A quick glance at the price charts of some market assets reveals periods of directional price movement and periods of price swings that reflect traders' decisions on whether to buy, sell, or exit the market. Traders are guided by logic and psychology in their decisions. The delicate balance between these two factors determines traders' preferences, which guide their decisions. Preferences, in turn, are shaped by market information, which is either analyzed rationally or distorted by psychological biases. This analysis provides meaning to the information that serves as a basis for further action. The purpose of the paper is to study the mechanisms that control traders' preferences, the formation of financial time series and the ability to predict future movements in asset prices. There are various approaches to studying market dynamics in order to predict future changes. One idea behind estimating the price movements of market assets is that past price values can indicate their future values according to observed market trends. It is the cornerstone of technical analysis (Miner, 2002). Elliott wave patterns (Elliott, 1994) and cycles, some of which may be related to business cycles, are used in market timing strategy (Millard, 1999).

Another approach argues that future price values have no relation to past price values, so asset prices resemble the movement of molecules (Osborn, 1959; Osborn and Murphy, 1984, cf. Fama, 1965; Malkiel, 1973). The efficient market hypothesis (EMH) (Fama, 1970) and related theories (e.g. Black and Scholes, 1973; Merton, 1973; Gulko, 1997, etc.) assume that asset prices reflect all available information, so future prices can not be predicted. However, there are also examples of predictable price behavior (for example, Working, 1960; Cowles and Jones, 1937; Kendall, 1953), which are considered anomalous under the random walk hypothesis. In addition, machine learning-based research has been reported to provide high forecasting accuracy for financial time series (e.g., Chang *et al*., 2009; Bitwai and Cohn, 2014; Patel *et al*., 2015; Hsua *et al*., 2016).

Behavioral economics, which focuses on the various psychological factors that influence people's economic decisions, takes a different approach from EMH (e.g., Kahneman and Tversky, 1979; Thaler, 1980, 1985; Banerjee, 1992). It states that the decisions of economic agents can largely be considered irrational and driven by psychological factors, so the hypothesis of rational agents can no longer be supported. A number of publications examine the mechanisms that determine traders decisions in financial markets (e.g. Shiller, 1981; Statman, 1995; Olsen, 1998; Barber and Odean, 1999). Because deviations from full rationality are systematic and can be modeled and studied, knowledge of these mechanisms can be used to improve predictions of future traders' behavior and market asset price movements. For example, in the case of herd behavior, market trends that have formed to date can be extrapolated to some future period. This indicates that the market is somewhat predictable and past prices can sometimes be used to predict the direction of future price changes (Lo and Mackinlay, 2002).

The adaptive market hypothesis (Lo, 2004, 2005) views the EMH and the behavioral approach as opposite sides of the same coin, interpreting the behavioral biases of market participants in an evolutionary perspective, as adaptation to changing conditions of the market environment, so that each group of market participants behaves accordingly. In its evolutionary aspect, the adaptive market hypothesis is close to the complex systems approach, which considers the market as a complex developing system of connected networks of interacting agents. Complexity theory, as applied to financial markets, examines nonlinearity, self-organization, emergent dynamics, and develops predictive models such as stock market crashes (Sornette, 2003).

While behavioral finance primarily focuses on individual-level biases, social finance—a new paradigm in financial research—focuses on cultural characteristics, including information signals, beliefs, strategies, etc., adopted by larger groups of investors. “Analysis of social interactions promises to provide greater insight into where heuristics come from, and to offer a foundation for understanding shifts in investor sentiment. As such, it can potentially offer a deeper basis for understanding the causes and consequences of financial bubbles and crises” (Hirshleifer, 2015, p. 44).

A common assumption in the approaches mentioned is that market asset prices reflect investors' reactions to information. But it is necessary to ask questions: 1) how investors process information; 2) whether the same information evokes the same meaning among different groups of investors; 3) does the way investors process information play a more significant role than the information itself? The purpose of the article is to find answers to these questions.

The paper proposes a new approach based on the processing of information by heterogeneous groups of investors that determines their behavior. Upon receipt, the information is processed, i.e. supplied with meaning. Different groups of investors may process the same information differently, providing different criteria by which to make sense of the information. These criteria can be defined as the selection environments in terms of specific coding rules (or sets of communication codes). Coding rules govern latent structures that organize different meanings into structural components (Leydesdorff, 2010). “Meanings originate from communications and feedback on communications. When selections can operate upon one another, a complex and potentially non-linear dynamics is generated” (Leydesdorff, 2021, p. 15). Meanings give rise to expectations regarding possible states of the system, which are generated with respect to the future. Expectations operate as feedback on the current state (i.e. against the arrow of time). In other words, the system simultaneously takes into account its past, present and future state, in line with Bachelier's observation that “past, present and even discounted future events are reflected in the market price” (Bachelier, 1964). Expectations provide a source of additional options for possible future states of the system that are available but not yet realized.

The more options a system has, the greater the likelihood that the system will deviate from its previous state in the process of autocatalytic self-organization. A measure of additional capacity is redundancy, which is defined as the addition of information to its maximum informational content (Brooks and Wiley, 1986). Regarding innovation studies, redundancy is associated with synergy in the Triple Helix (TH) model of university-industry-government relations (e.g. Etzkowitz and Leydesdorff, 1995, 1998; Leydesdorff, 2003; Park and Leydesdorff, 2010; Leydesdorff and Strand, 2013, etc.). The interaction between differently shaped traders' expectations can ultimately lead to nonlinear market price dynamics.

The paper's contribution is two-fold: 1) in a narrow sense, it presents a model of market dynamics that sheds light on the mechanisms that govern the formation of prices of market assets and, in some cases, allows us to predict future price movements; 2) in a broader sense, it is a step forward in the development of a theory of meaning, which brings this theory into a wider practical area. The first research objective of this paper is to test the applicability of the general concept of information and meaning to describe market price dynamics. The second goal of the study is to provide a quantitative description of market price movements based on the evolutionary dynamics of traders' expectations. This dynamics is captured by a nonlinear evolutionary differential equation, the solutions of which provide recognizable patterns. One can use these patterns to predict future asset price movements, which allows make more accurate forecasts of future price changes.

## II. Literature review

The approach that uses meaning generation in inter-social communications is not rooted in financial literature. However, models with similar basic characteristics can be found in the literature. Financial time series modeling has been a subject of debate among scholars and practitioners. A big branch of the literature is devoted to machine forecasting of time series. Related approaches include traditional Machine Learning (ML) models (e.g. Bahrammirzaee, 2010; Mullainathan, S. and Spiess, 2017, etc.) and emerging within the ML field Deep Learning (DL) models such as Artificial Neural Networks (ANN ), Recurrent Neural Network (RNN), Convolutional Neural Network (CNN), and Deep Multilayer Perceptron (DMLP) (e.g., Schmidhuber, 2015; Sokolov *et al*., 2020). The advantage of the analytical approach compared to the numerical one is that in analytical models it is possible to trace the properties of time series depending on the behavior of investors. Frequently observed and difficult to explain facts such as excess volatility (LeRoy & Porter, 1981; Shiller, 1981), volatility clustering (Mandelbrot, 1997), etc., have led to a growing body of literature on heterogeneous agent based models (HAM) (e.g. Hommes, 2006, Chiarella, Dieci and He, 2009), in which the financial market includes different groups of agents. Agent interactions can generate complex market dynamics involving both chaos and stability (e.g., Brock and Hommes, 1997; Lux, 1995; Bonabeau, 2002). Kaizoji (2004) showed that periodic chaos in asset price dynamics can be observed in a simple model of financial markets with two groups of agents, which can be explained by the heterogeneity of traders' trading strategies. The behavior of financial markets during periods of crisis can be described as rogue waves (e.g., Jenks, 2020). This type of waves is analytically depicted in a nonlinear option pricing model (Yan, 2010) and can be used to describe the possible mechanisms of the rogue wave phenomenon in financial markets. Rogue waves can be detected in Korteweg de Vries (KdV) systems when real nonintegrable effects, higher order nonlinearity, and nonlinear diffusion are taken into account (Lou and Lin 2018). Dhesi and Ausloos (2016), while studying the behavior of agents reacting to time-dependent on the log returns within an Irrational Fractional Brownian motion model, observed a kink-like effect reminiscent of soliton behavior. Next, they posed the question: what is the differential equation whose solution describes this effect? In some ways, the idea behind this approach is similar to that used in opinion dynamics models (e.g., Zha *et al*., 2020; Granha *et al*., 2022). Binge and Boshoff (2021) introduced a pseudo-resales method for constructing art prices. They proposed a regression-based method that provides a more accurate estimate of price changes and allows the study of bubbles in a range of asset markets. The gap between an asset's price and fundamental value will exhibit temporary explosive behavior. The resulting bubble is expected to eventually burst and hence a trend reversal can be predicted. In the absence of structural changes in fundamentals, the period of explosive prices has a non-fundamental explanation, namely cognitive bias. The inability of traders to properly process available information can cause systematic patterns in price movements, such as seasonality. This inability can be explained by a cognitive bias on the part of investors (Hirshleifer *et al*., 2020; Fang *et al*., 2021), who use ease-of processing heuristics when processing information relevant to the pricing of market assets. Knowledge of seasonality mechanisms allows us to predict the dynamics of repo rates. Biased beliefs about future price movements are an important driver of market prices. Changes in traders' expectations about future stock market returns can explain facts about stock market price movements (De la O and Myers, 2021). Jin and Sui (2022) showed that investors' beliefs about future market returns can cause excessive volatility in stock market returns. These beliefs depend primarily on recent past returns and can provide significant returns predictability. The behavior of a group of traders in interpreting external information also influences the movement of stock prices. Massa *et al*. (2022) showed that the reallocation of strategic risks of business groups in response to news that can be interpreted as “bad” information is an additional factor in shaping stock returns. Duffy *et al*. (2021) studied how exchange traded funds (ETFs) affect asset prices and turnover in a laboratory asset market consisting of three subgroups of traders, with each subgroup employing its own preferential trading strategy. The three-part market structure is also described within the SIR model, which is used to explain the spread of investor ideas and bubbles (Shiller, 2019). SIR-related models can explain price jumps, bubbles, and the serial correlation structure in asset returns with different lags (Burnside *et al*., 2016; Hirshleifer, 2020). The SIR model, in turn, can be coupled with the TH model and reveal the corresponding nonlinear dynamics described by the KdV equation (Ivanova, 2022). Noussair and Popescu (2021) provide evidence that market asset co-movements in the absence of a common fundamentals shock can be explained by behavioral factors driven by asymmetric information between informed and uninformed traders. A new paradigm for understanding financial markets is social finance (Akçay and Hirshleifer, 2021). Social finance argues that the behavior of market participants depends on the accepted cultural characteristics that determine the operating patterns of markets. These traits are common to larger investor groups and provide bias in the transmission of information between investor groups. Investors' financial characteristics are volatile and may change over time.

## III. Method

The paper's method draws on the general theory of meaning in inter-social communications (Leydesdorff, 2021), which is based on an information-theoretic approach to measuring synergy in the Triple Helix model of innovation.The market is a complex social system comprised of various groups of traders – pension funds, banks, hedge funds, individuals, etc. whose compound dynamics drive market price movement. Traders behave in accordance with their investment preferences. Instead of using an approach to market microstructure that relies on computational agent-based models that simulate the behavior of individual traders (e.g., O'Hara, 1995), one can focus on information processing among larger groups of investors.With respect to traders' sentiments they can be subdivided into momentum, reversal, and interactive traders. Traders make their decisions based on communicated information. However this information should first be given meaning (as a 'signal') or discarded (as noise). Each group uses different criteria for filtering information in accordance with their behavioral biases. That is the same information is considered from different perspectives (or positions), and can be supplied with different meanings. Mechanisms of information and meaning processing are different. Whereas information can be communicated via a network of relationships meanings are provided from different positions (Burt, 1982). Meaning cannot be communicated but only shared when the positions overlap. The processing of meaning can enlarge or reduce the number of options that can be measured as redundancy (Leydesdorff and Ivanova, 2014). The calculus of redundancy is complementary to calculus of information. Shannon (1948) defined information as probabilistic entropy: $H = -\sum pi\ i\ logpi$ which is always positive and adds to the uncertainty (Krippendorff, 2009). One can consider two overlapping distributions with information contents $H1$ and $H2$ (Figure 1)

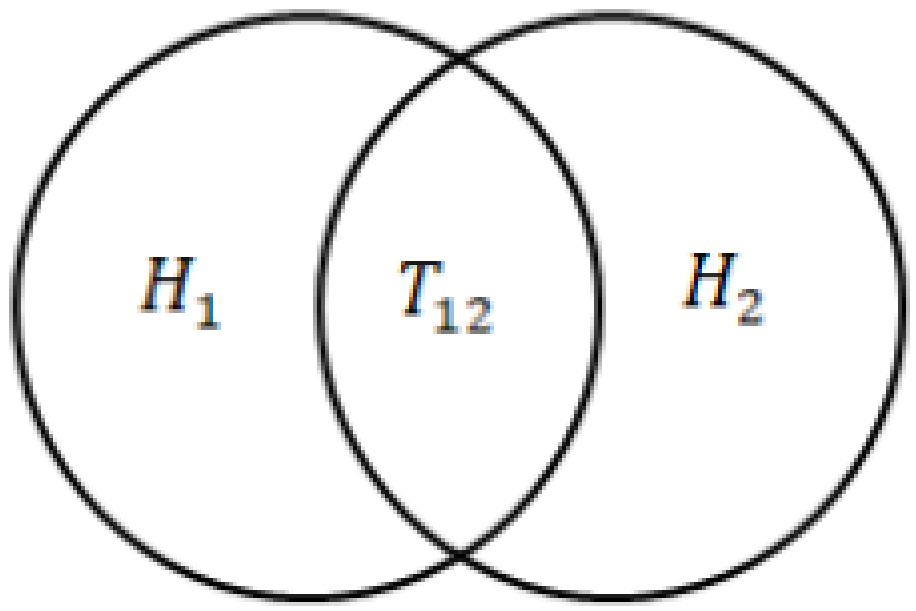

**Figure 1**: Set-theoretical representation of two overlapping distributions with informational contents $H_1$ and $H_2$

Total distribution is the sum of two distributions minus overlapping area, since it is counted twice:

$$H_{12} = H_1 + H_2 - T_{12} \tag{1}$$

Overlapping area relates to mutual or configurational (McGill, 1954) information $T_{12}$. The formula linking cummulative distribution ($H_{12}$) with $H_1$, $H_2$ and $T_{12}$ is:

$$T_{12} = H_1 + H_2 - H_{12} \quad (2)$$

Analogously configurational information in three dimensions (e.g., Abramson, 1963) is:

$$T_{123} = H_1 + H_2 + H_3 - H_{12} - H_{13} - H_{23} + H_{123} \quad (3)$$

Figure 2 is a set-theoretical representation of three overlapping distributions.

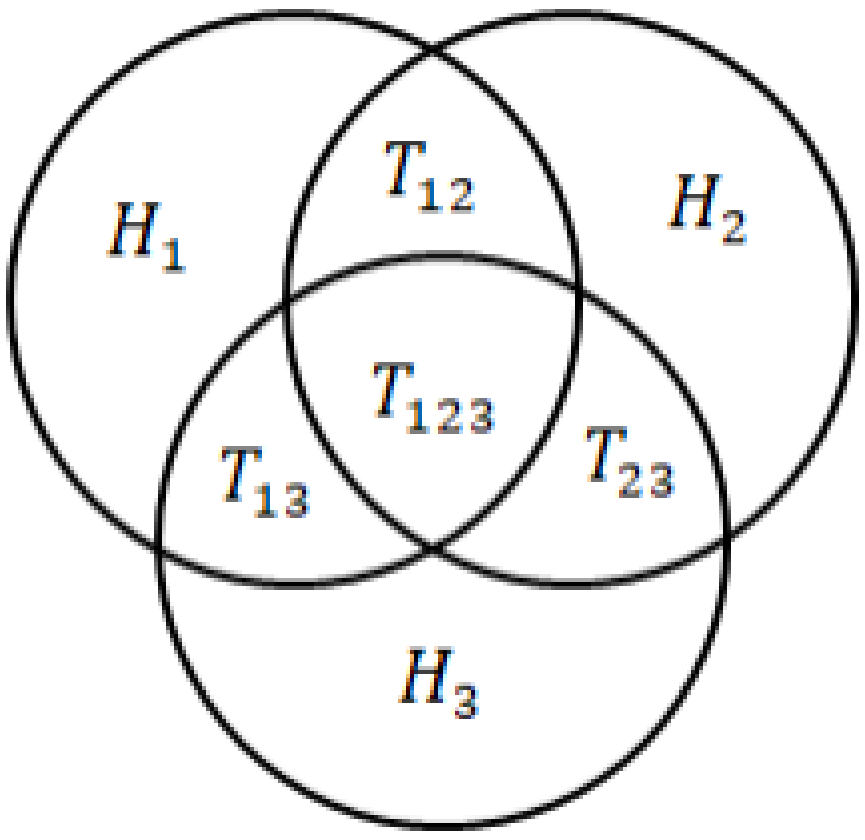

**Figure 2:** Set-theoretical representation of three overlapping distributions with informational contents $H_1$, $H_2$ and $H_3$

However $T_{123}$ is no longer Shannon-type information since it is negative.[2] The sign alters with each newly added distribution (e.g. Krippendorff, 2009). Technically the problem of sign change can be solved by introducing "positive overlapping" (Leydesdorff & Ivanova, 2014). This time one does not correct for overlap which is counted twice and therefore redundant, but assume other mechanisms with which two distributions influence one another. These mechanisms are different from the relational exchange of information and lead to an increase of redundancy so that overlapping area is added instead of being subtracted. In formula format, this is represented as follows:

$$H_{12} = H_1 + H_2 + R_{12} \quad (4)$$

[2] This is true for the configuration shown in Figure 3. In general the resulting value can be positive, negative or zero depending upon the relative sizes of terms.

It follows that $R_{12}$ is negative ($R_{12} = -T_{12}$) and is hence a redundancy - reduction of uncertainty[3] (analogously: $R_{123} = T_{123}$). That is in measuring configurational information in three (or more) dimensions, one measures not Shannon-type information but mutual redundancy. Since this measure provides a negative amount of information it can be considered an indicator of synergy among three sources of variance (Leydesdorff, 2008b).

Three groups of traders communicate via a network of relations. But there is another mechanism on the top of the structural network which drives the system evolution. Communicated information is processed differently by each group of traders in accordance with different sets of coding rules (communication codes) so that groups are differentiated with respect to their positions toward processing the information. For example, the same information can be supplied with different meaning and, according to traders' sentiments, can be accepted as a signal to buy or sell by different agents. The sets of communication codes are latent but can be partly correlated by forming a correlation network on the top of the relational one. The generation of meaning is provided from the perspective of hindsight. The structural differences among the coding and decoding algorithms provide a source of additional options in reflexive and anticipatory communications, meaning that the generating structures act as selection environments (Leydesdorff, 2021). These additional options are the source of variations. Variations arise when market eventually changes from its previous state. When a system comprises three or more groups (agents) each third agent disturbs interaction between the other two.  This mechanism is known as “triadic closure” and drives the system’s evolutionary dynamics (Granovetter, 1973; Bianconi *et al.*,  2014, de Nooy and Leydesdorff, 2015). Simmel (1902) pointed toward qualitative difference between dyads and triads. Triads can be transitive or cyclic (Batagelj *et al*., 2014). Two cycles may emerge – positive (autocatalitic) and negative (stabilizing) ones (Figure 3). Autocataitic cycle reinforce the change from the previous system state (the system self-organizes) while the stabilising cycle keeps the system from transformation.

---

[3] Redundancy is defined as a fraction of uncertainty that is not used for the uncertainty $H$ that prevails: $R = 1 - \frac{H}{H_{max}} = \frac{H_{max} - H}{H_{max}}$. $H_{max}$ – maximum possible entropy. Adding new options increases $H_{max}$ and redundancy.

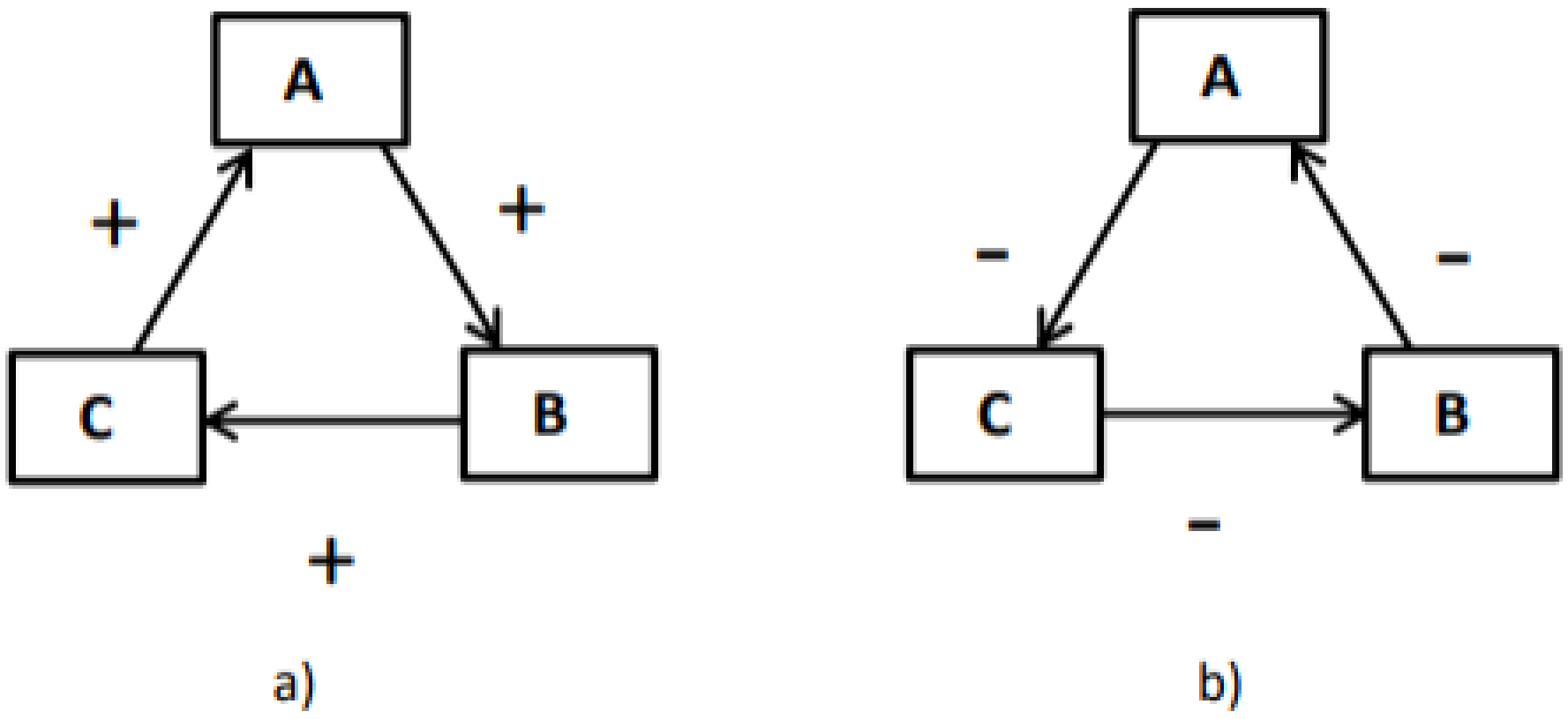

**Figure 3**:  Schematic of three-component positive a) and negative b) cycles (Adapted from Ulanovitz, 2009)

The dynamics of information and meaning can be evaluated empirically using the sign of mutual information ($R$) as an indicator. The balance between stabilization and self-organization can be simulated in terms of the rotations of the two vectors $P$ and $Q$ (Ivanova and Leydesdorff, 2014a, 2014b)

$$R \sim P^2 - Q^2 \qquad (5)$$

The first term in Equation 5 is due to historical realization and adds to positive entropy and the second term corresponds to self-organization and augments negative entropy. Historical realization relates to historically realized options which are generated via a recursive mode while self-organization bears on new, not yet realized options, generated via an incursive mode.[4]
Comparing two expressions for redundancy provided by Equations 1 and 2 one can identify positive and negative terms in Equation 1 with positive and negative terms in Equation 2.
The trade-off between historical realization and self-organization (Equation 5) leads to redundancy cyclical evolution (Ivanova and Leydesdorff, 2014b). Dubois shows (2019) that for temporal cyclic systems probabilities $p_i$ oscillate around their average values $p_{i0}$ in a harmonic mode. Non-harmonic mode is also possible (Ivanova, 2022).
For non-harmonic oscillations, one can derive (Appendix A):

$$\frac{1}{k}\frac{d^2p_i}{dt^2} = -(p_i - p_{i0}) + \alpha(p_i - p_{i0})^2 + C_i \qquad (6)$$

The probability density function $P$ satisfies the following non-linear evolutionary equation (Appendix B):

$$P_T + 6PP_X + P_{XXX} + C_1 = 0 \qquad (7)$$

which is the generalization of the Korteweg-de Vries equation:

$$U_T + UU_X + U_{XXX} = 0 \qquad (8)$$

Equation 7 admits soliton solutions. A single soliton solution is:

$$P(X,T) = 2\left(\frac{\kappa}{2}\right)^2 ch^{-2}\left[\frac{\kappa}{2}\left(X - 4\left(\frac{\kappa}{2}\right)^2 T + \frac{C_1}{2}T^2\right)\right] - C_1T \qquad (9)$$

Setting $\rho = \frac{\kappa}{2}$, Equation 9 can be written in a more general form:

$$P(X,T) = n(n+1)\rho^2 ch^{-2}\left[\rho\left(X - 4\rho^2 T + \frac{C_1}{2}T^2\right)\right] - C_1T \quad (10)$$

[4] Recursive systems use their past states to modulate the present ones, incursive (or anticipatory) systems employ possible future states to shape their present states (e.g. Rosen, 1985; Dubois, 1998; Leydesdorff & Dubois, 2004)

An impulse in the form of Equation 10 eventually evolves in a train of *n* solitary waves with amplitudes: $2\kappa^2$, $8\kappa^2$, $18\kappa^2$ ...$2n^2\kappa^2$ and the corresponding velocities $4\kappa^2$, $16\kappa^2$, $32\kappa^2$, ... $4n^2\kappa^2$[5] **(**Miura, 1976**).**

There is also a direct method for finding soliton solutions for Equation 8 which incorporate multiple solitons with arbitrary amplitudes so that the $N$ – soliton solution takes the form:

$$P = 2\frac{d^2}{dX^2}logF_N \qquad (11)$$

where:

$$F_N = \sum_{\mu=0,1} exp\left(\sum_{i=1}^{N} \mu_i\eta_i + \sum_{1\le i<j}^{N} \mu_i\mu_j A_{ij}\right) \qquad (12)$$

Here $\eta_i = k_iX - k_i^3T$; $A_{ij}$ - phase shifts of the solitons: $e^{\,A_{ij}} = \left(\frac{k_i-k_j}{k_i+k_j}\right)^2$ (Ablowitz and Segur, 1981)[6]. It follows from Equation 12 that the corresponding N-soliton solution for Equation 8 is:

$$\Phi_N = exp\left[-\frac{C}{2}tx^2 + Ax + B\right] \cdot \sum_{\mu=0,1} exp\left(\sum_{i=1}^{N} \mu_i\eta_i + \sum_{1\le i<j}^{N} \mu_i\mu_j A_{ij}\right) \qquad (13)$$

The additional term at the right hand side of Equation 10 corrects the amplitude of the solitons with the lapse of time. Wave, described by Equation 9, moves to the right and after time span: $T_1 = {8k^2}/{C_1}$ returns to the origin. In case of a train of solitons there is a relationship between soliton amplitudes and time intervals: $\frac{A_i-A_j}{T_i-T_j} = const$. There are also periodic solutions of Equation 8 (Appendix C, cf. Lax, 1974).

Information obtained via informational exchange is processed with communication codes and expectations with respect to future time are generated at a system's level. These expectations

---

[5] Equation 12 refers to a net soliton solution. In the case of an arbitrary initial distribution it evolves in a train of solitons moving off to the right and an oscillatory dispersive state moving off to the left.

[6] The sum over $\mu = 0,1$ refers to each of the $\mu_i$. E.g. performing the calculation for N=3 yields $F_3 = 1 + e^{\eta_1} + e^{\eta_2} + e^{\eta_3} + e^{\eta_1+\eta_2+A_{12}} + e^{\eta_1+\eta_3+A_{13}} + e^{\eta_2+\eta_3+A_{23}} + e^{\eta_1+\eta_2+\eta_3+A_{12}+A_{13}+A_{23}}$

present non-realized but possible options, distributed along time interval, which can trigger subsequent actions. Here expectations are analytical events (options) and actions are historical events,[7] which can be observed over some time as a response to the expectations (as if expectations move against the arrow of time and turn into actions). In other words, there is a dynamic of the actions in historical events at the bottom and a dynamic of expectations at the upper level operating reflexively. Expectations are eventually transformed into actions and represent new system states.[8]

Initial expectations arise as a set of beliefs with respect to past, present and future states of market[9]. These expectations are projected to the future and further stratified following the non-linear dynamics provided by Eq. 7 as a train of solitons. Finally expectations turn into traders' actions driving market asset price changes. This mechanism operates on all price and time scales generating a self-similar fractal structure.

## IV. Data

The paper relies on publicly available data sets. The data comprise Australian dollar daily data (2001.09.21 -2004.05.13), Swiss franc hourly data (2021.01.07 – 2021.04.12), natural gas daily data (2016.03.02 – 2017.02.13), Corn daily data (2020.08.10 -2021.07.23), and S&P500 index monthly data (2009.02.01 – 2022.01.01). Various data sets are specially chosen in order to test the model applicability across different markets and different time scales. I compare the model results with the empirically observed data on FX, energy, food, and indices markets. Since the S&P 500 index market data take a fairly long time interval I use inflation-adjusted data.

---

[7] Shannon (1948) defined the proportion of non-realized but possible options as redundancy, and the proportion of realized options as the relative uncertainty or information.

[8] According to the second law of thermodynamics a system's entropy increase with time. For isolated systems it can reach thermodynamic equilibrium

[9] The set of such states represent phase space

## V. Results

It follows from the model that when non-linear effects prevail (Equation A 10) asset prices develop according to trends that can be described by the non-linear evolutionary equation (7). In other cases, there may be other mechanisms that determine price dynamics.

To answer the question of whether the concept of information and meaning communication can be applied to describe the dynamics of market prices, one can try to find patterns that can be described by the model. I compare the model's results with empirically observed data from foreign exchange, energy, food, and indices markets. The same patterns can be found across different markets and (due to the fractal nature of stock and financial markets) across different time frames. The selection of different assets from different markets (FX, energy, food, and indices) is made to illustrate the application of the model to different markets. Because market data for the S&P 500 index takes a fairly long time frame I used inflation-adjusted data.

Figure. 4 shows time series for a) Australian dollar (AUD/USD), b) Swiss franc (USD/CHF), c) Natural gas (NGas), d) Gold (XAU) and e) Standard and Poor's 500 stock market index (S&P500) and their model approximation (*f*). The data are approximated by one or two series of solitons obtained from Equation. 13. The approximation parameters are given in Table 1. Here $A_i$–amplitudes of the solitons ($i$ - is the number of solitons in the series), $k_i$ – are the parameters $k$, $T_i$ – time intervals (Equation 12), β - vertical shift (are introduced in order to equate the beginning of the first waves to zero).

| | AUD/USD | USD/CHF | NG | Corn | S&P 500 |
|---|---|---|---|---|---|
| β | 0.4848 | 0.878 | 1.627 | 310.75 | 966 |
| $A_1$ | 0.014 | 0.016 | 0.44 | 71.75 | 473.9 |
| $(k_1; T_1)$ | (0.043; -29.28) | (0.005; -350) | (0.026; 42.1) | (0.03; -54.16) | (0.062; -21.01) |
| $A_2$ | 0.01 | 0.046 | 0.88 | 208.21 | 1368 |
| $(k_2; T_2)$ | (0.03; -176.16) | (0.009; -1050) | (0.032; 100.3) | (0.04; -122.4) | (0.057; -66.34) |
| $A_3$ | 0.009 | 0.064 | 1.33 | 370.57 | 2191 |
| $(k_3; T_3)$ | (0.03; -418.1) | (0.011; -1450) | (0.032; 160.2) | (0.02; -201) | (0.053; -106.9) |
| $A_4$ | 0.01 | - | 1.77 | - | 3126 |
| $(k_4; T_4)$ | (0.03; -612.17) | - | (0.03; 218.6) | - | (0.048; 151.6) |

**Table 1** Approximation parameters for empirically observed data on market assets

Daily Australian dollar data (2001.09.21 -2004.05.13), daily natural gas data (2016.03.02 – 2017.02.13) and monthly inflation adjusted S&P500 data (2009.02.01 – 2022.01.01) were fitted with four solitons. Hourly Swiss franc data (2021.01.07 – 2021.04.12) and daily corn data (2020.08.10 -2021.07.23) were fitted with a series of three solitons. The approximation parameters satisfy the relationship predicted by the model, i.e. the slope of the upward trend for every two solitons doesn't change: $\frac{A_i - A_j}{T_i - T_j} = const.$

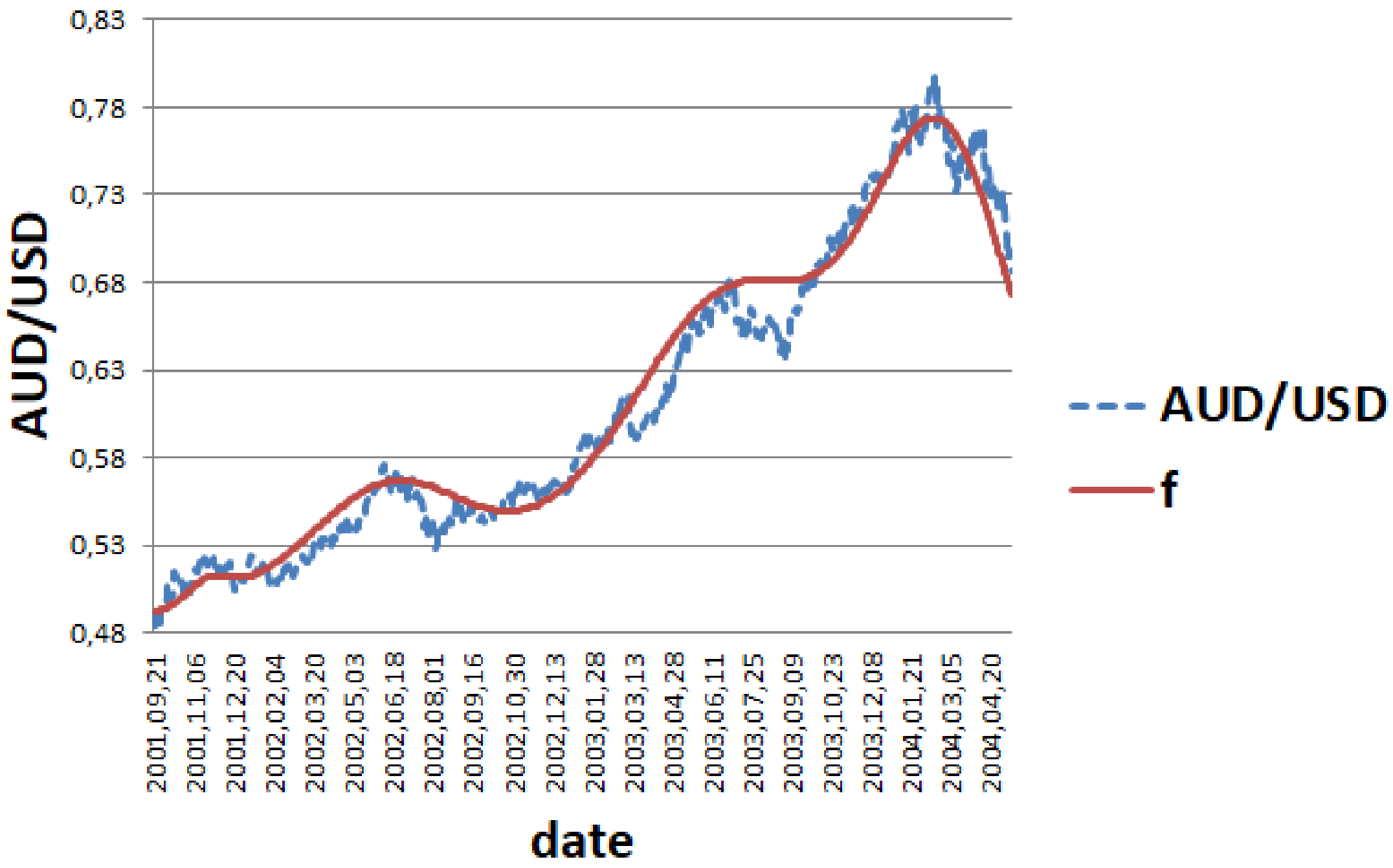
0,83
0,78
0,73
0,68
0,63
0,58
0,53
0,48
AUD/USD
2001,09,21
2001,11,06
2001,12,20
2002,02,04
2002,03,20
2002,05,03
2002,06,18
2002,08,01
2002,09,16
2002,10,30
2002,12,13
2003,01,28
2003,03,13
2003,04,28
2003,06,11
2003,07,25
2003,09,09
2003,10,23
2003,12,08
2004,01,21
2004,03,05
2004,04,20
date
AUD/USD
f

a)

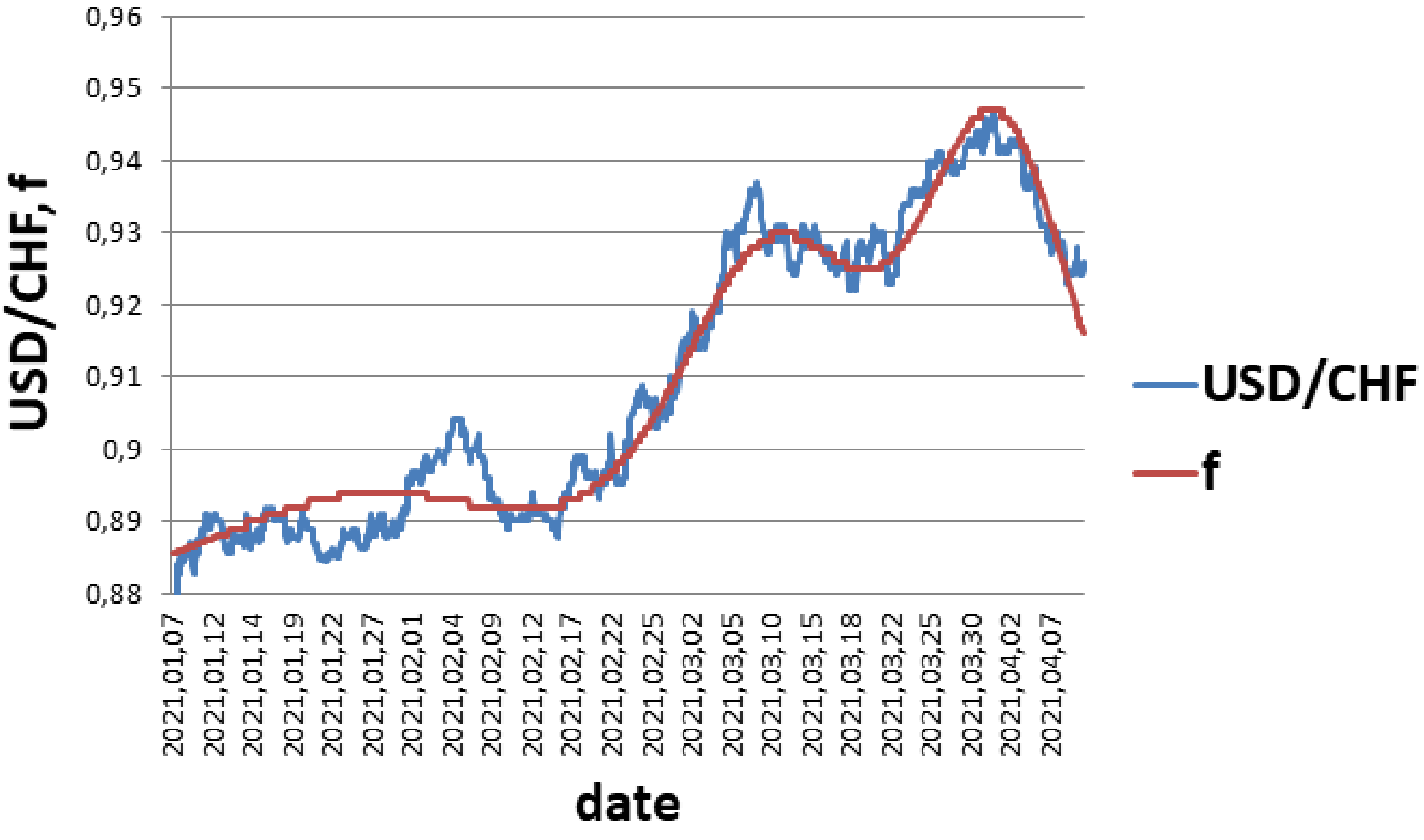
0,96
0,95
0,94
0,93
0,92
0,91
0,9
0,89
0,88
USD/CHF, f
2021,01,07
2021,01,12
2021,01,14
2021,01,19
2021,01,22
2021,01,27
2021,02,01
2021,02,04
2021,02,09
2021,02,12
2021,02,17
2021,02,22
2021,02,25
2021,03,02
2021,03,05
2021,03,10
2021,03,15
2021,03,18
2021,03,22
2021,03,25
2021,03,30
2021,04,02
2021,04,07
date
USD/CHF
f

b)

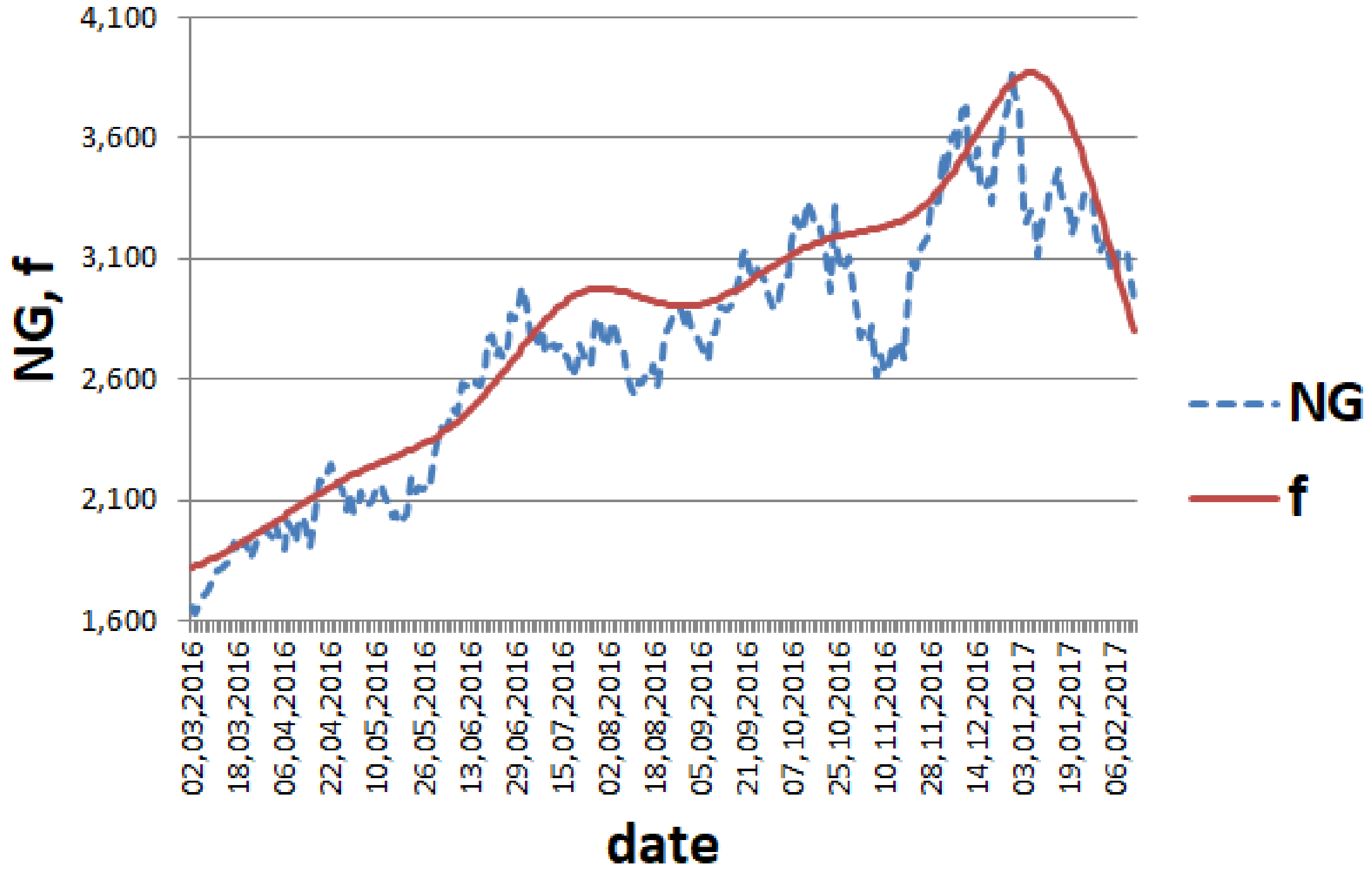

4,100
3,600
3,100
2,600
2,100
1,600
NG, f
02,03,2016
18,03,2016
06,04,2016
22,04,2016
10,05,2016
26,05,2016
13,06,2016
29,06,2016
15,07,2016
02,08,2016
18,08,2016
05,09,2016
21,09,2016
07,10,2016
25,10,2016
10,11,2016
28,11,2016
14,12,2016
03,01,2017
19,01,2017
06,02,2017
date
NG
f

c)

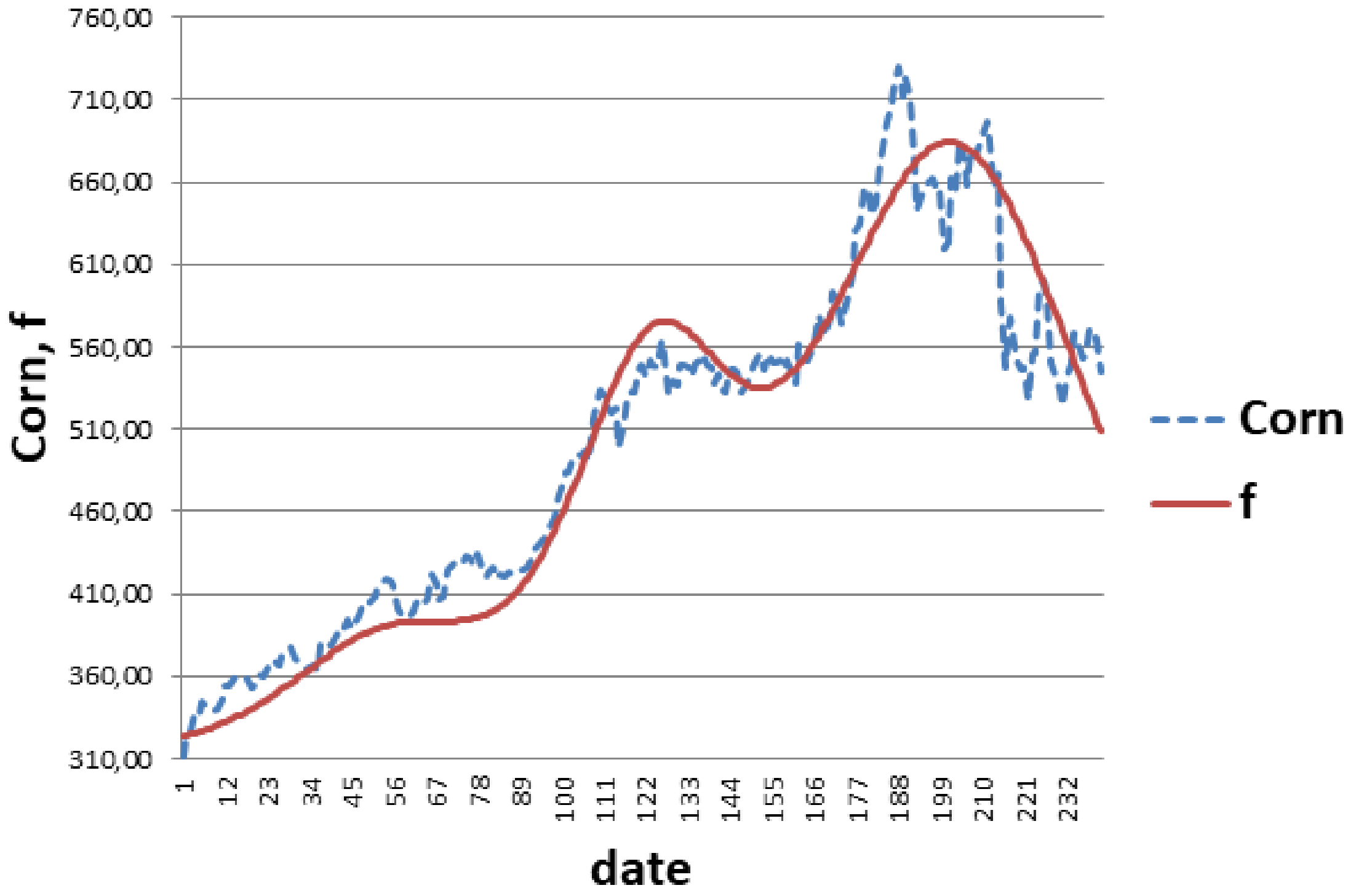

760,00
710,00
660,00
610,00
560,00
510,00
460,00
410,00
360,00
310,00
Corn, f
1
12
23
34
45
56
67
78
89
100
111
122
133
144
155
166
177
188
199
210
221
232
date
Corn
f

d)

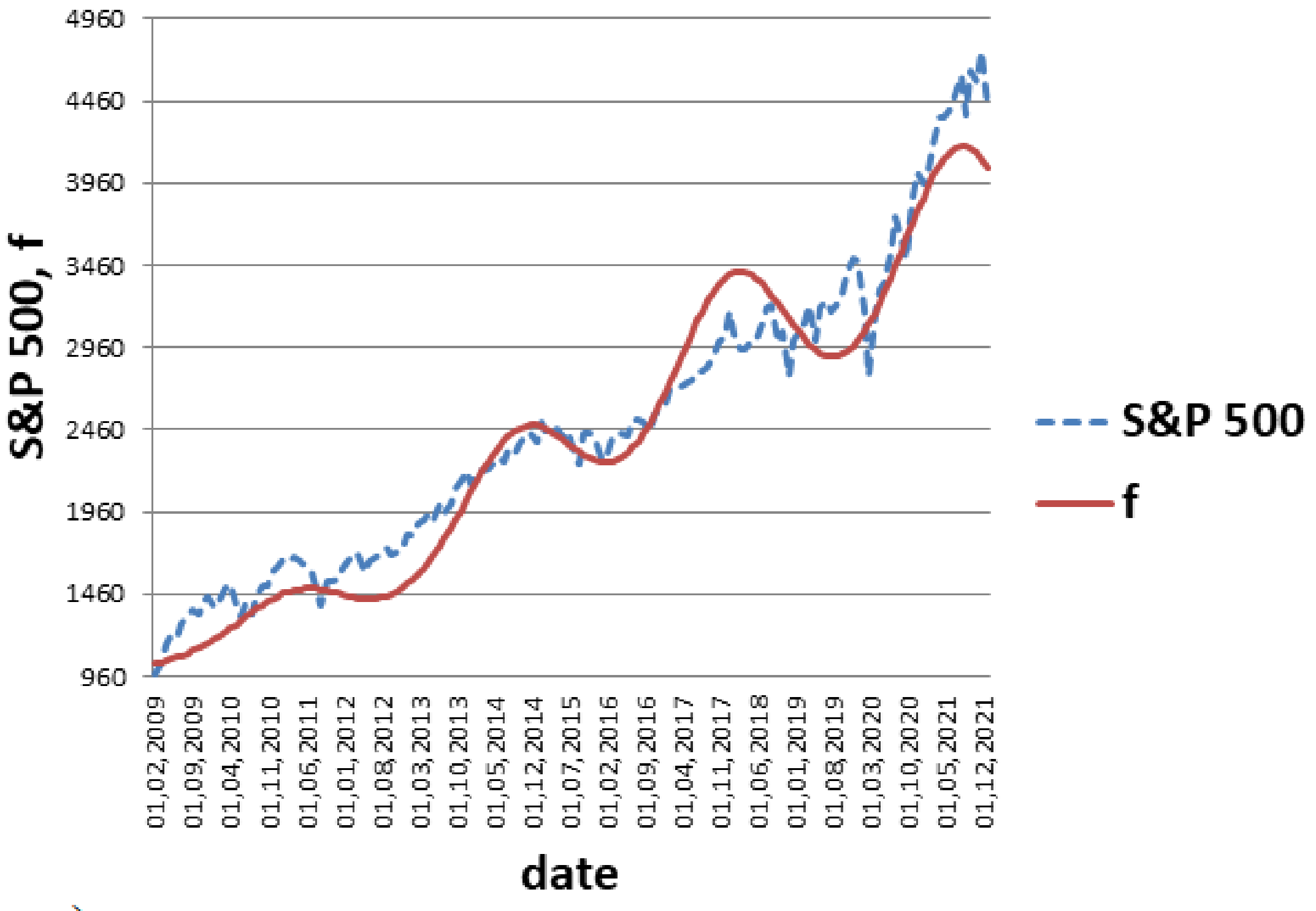

e)

**Figure 4:** Time series for the a) daily Australian dollar (AUD/USD), b) hourly Swiss franc (USD/CHF), c) daily Natural gas (NG), d) daily Corn data and e) monthly Standard and Poor's 500 stock market index (S&P500) data; *f* - model fit.

Figure 5 shows a scatter plot of a) Australian dollar (AUD/USD), b) Swiss franc (USD/CHF), natural gas d), corn and e) S&P 500 index values compared to model predicted values (*f*). A straight line is a linear regression. Regression parameters are shown in Table 2.

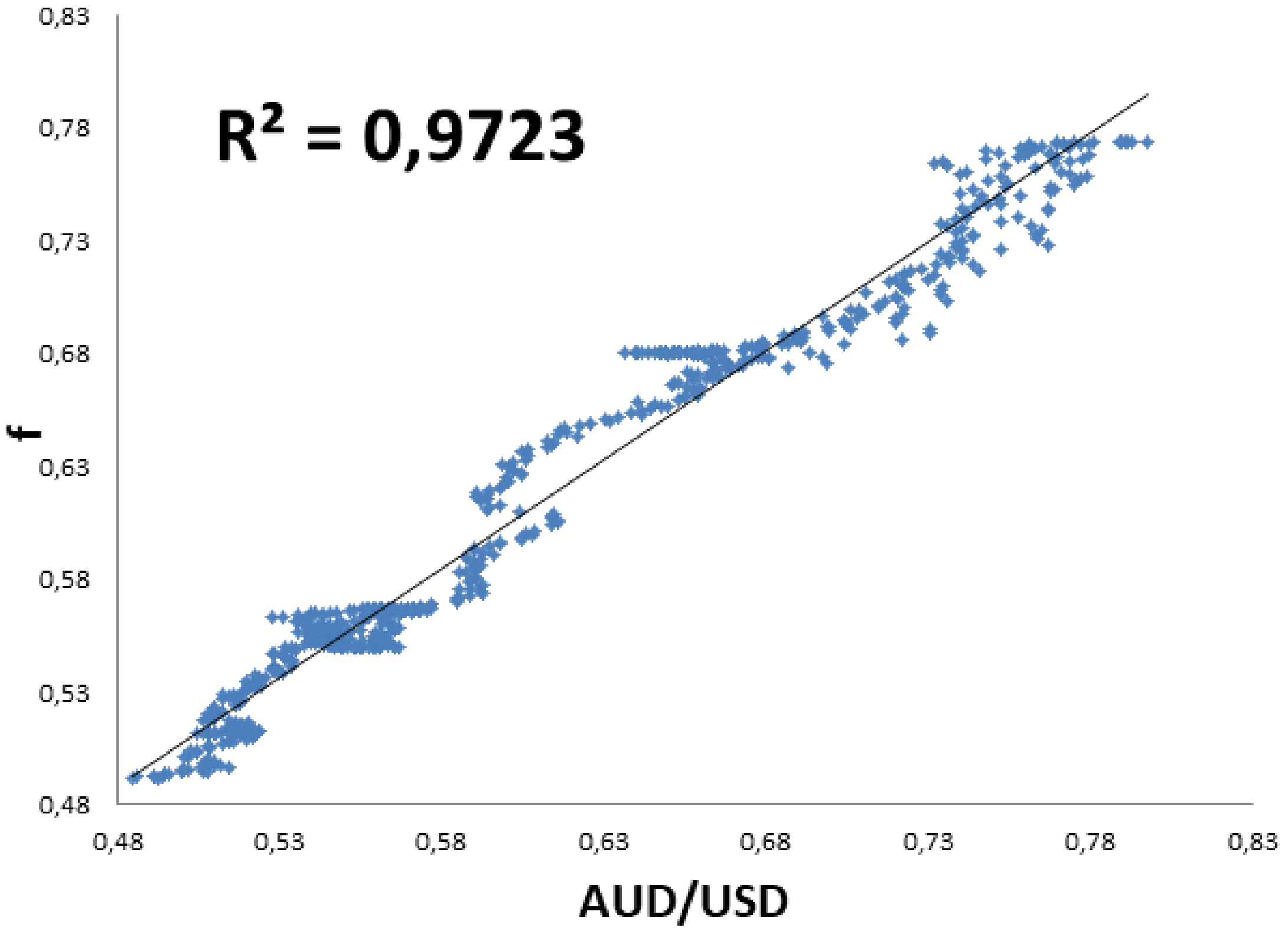

R² = 0,9723
f
0,83
0,78
0,73
0,68
0,63
0,58
0,53
0,48
0,48
0,53
0,58
0,63
0,68
0,73
0,78
0,83
AUD/USD

a)

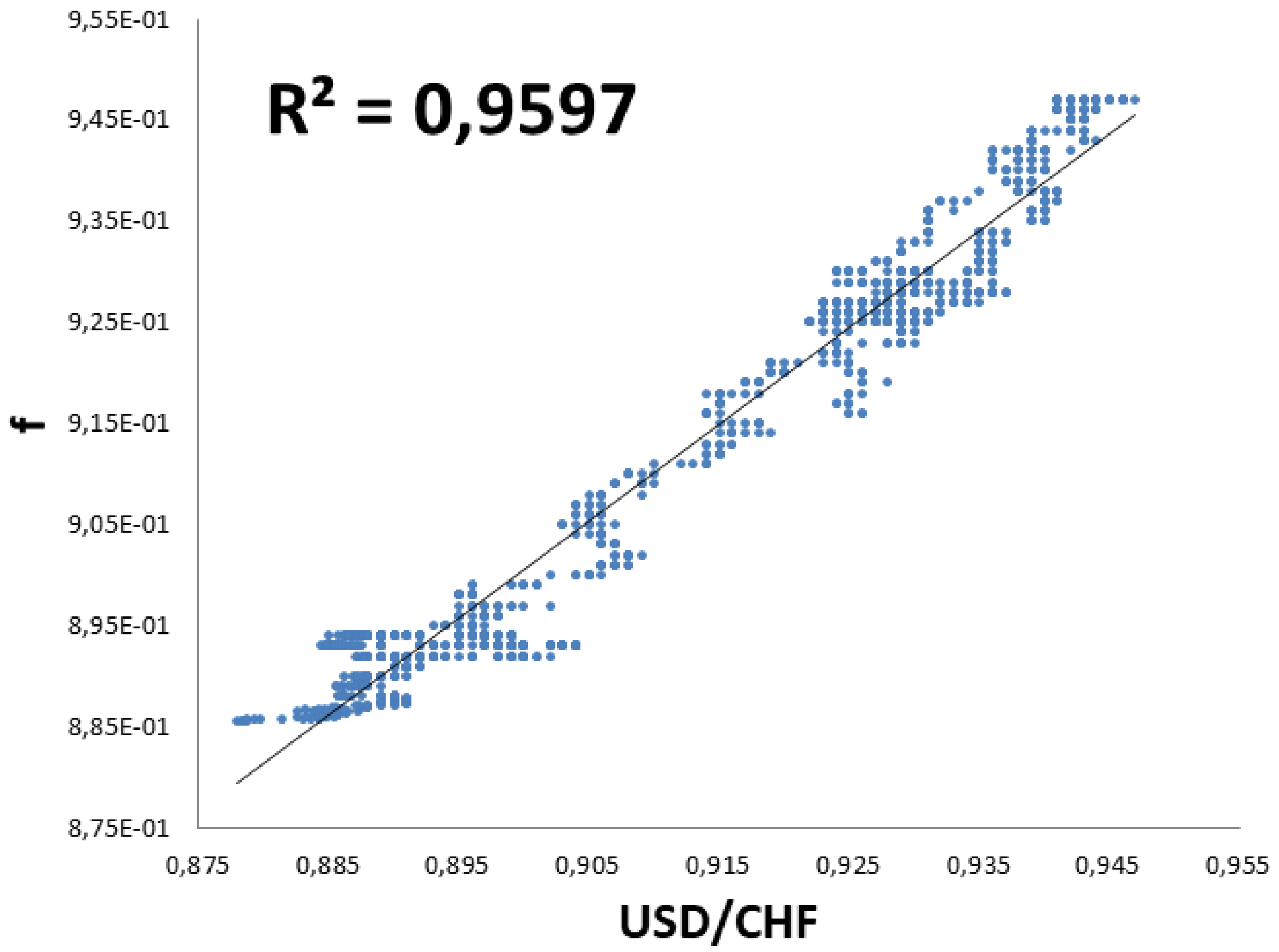

R² = 0,9597
f
9,55E-01
9,45E-01
9,35E-01
9,25E-01
9,15E-01
9,05E-01
8,95E-01
8,85E-01
8,75E-01
0,875
0,885
0,895
0,905
0,915
0,925
0,935
0,945
0,955
USD/CHF

b)

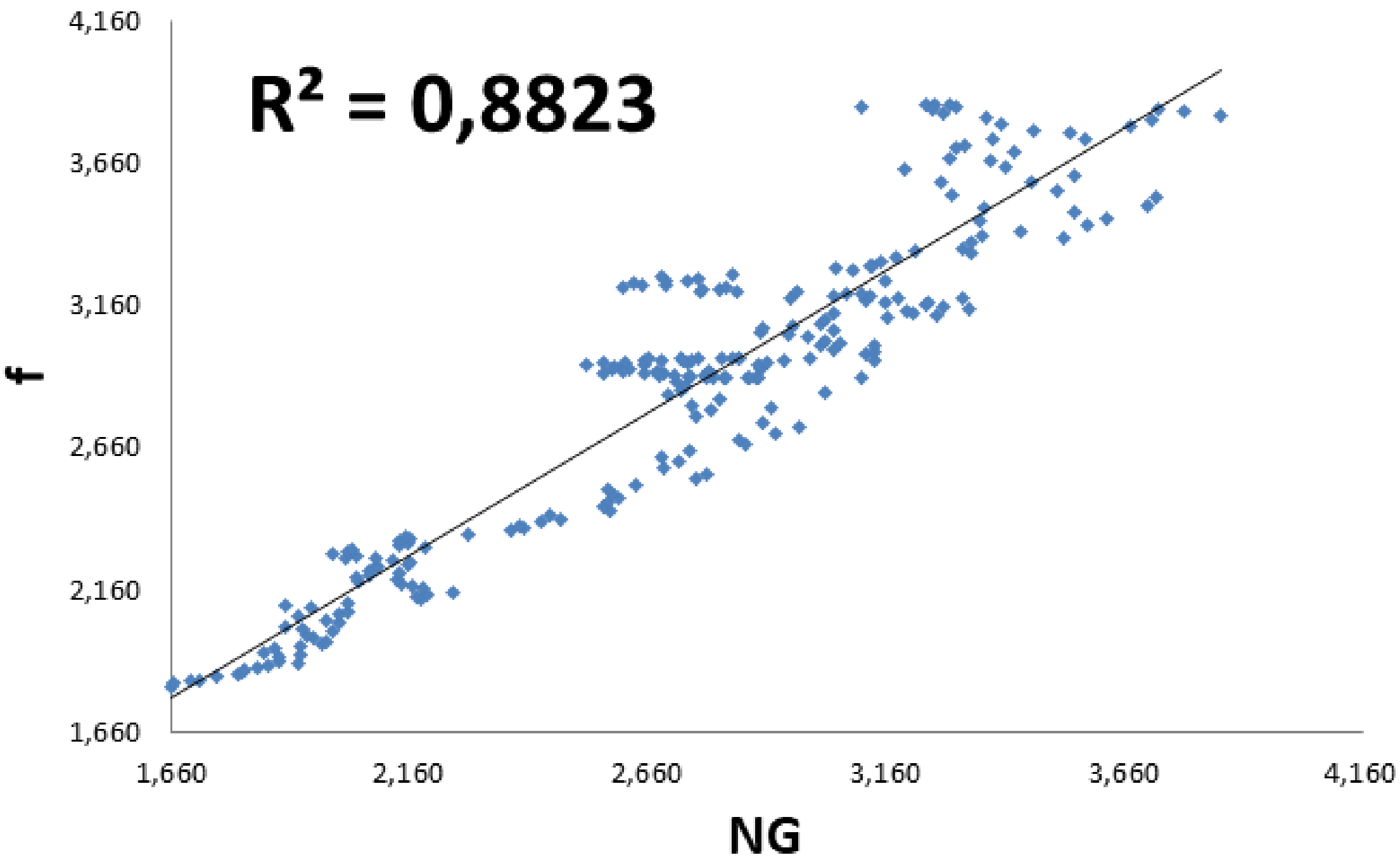
R² = 0,8823
4,160
3,660
3,160
2,660
2,160
1,660
f
1,660
2,160
2,660
3,160
3,660
4,160
NG

c)

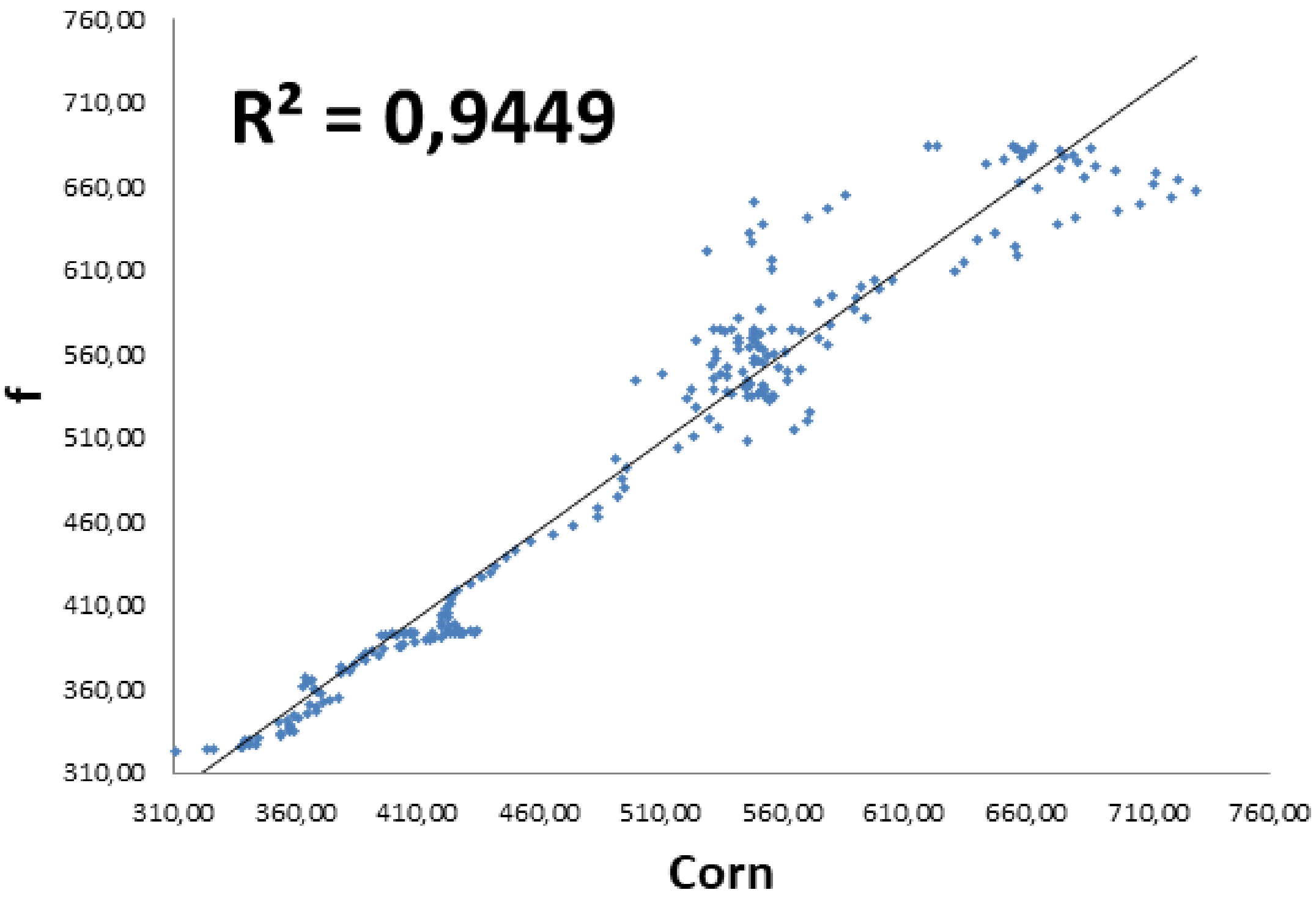
R² = 0,9449
760,00
710,00
660,00
610,00
560,00
510,00
460,00
410,00
360,00
310,00
f
310,00
360,00
410,00
460,00
510,00
560,00
610,00
660,00
710,00
760,00
Corn

d)

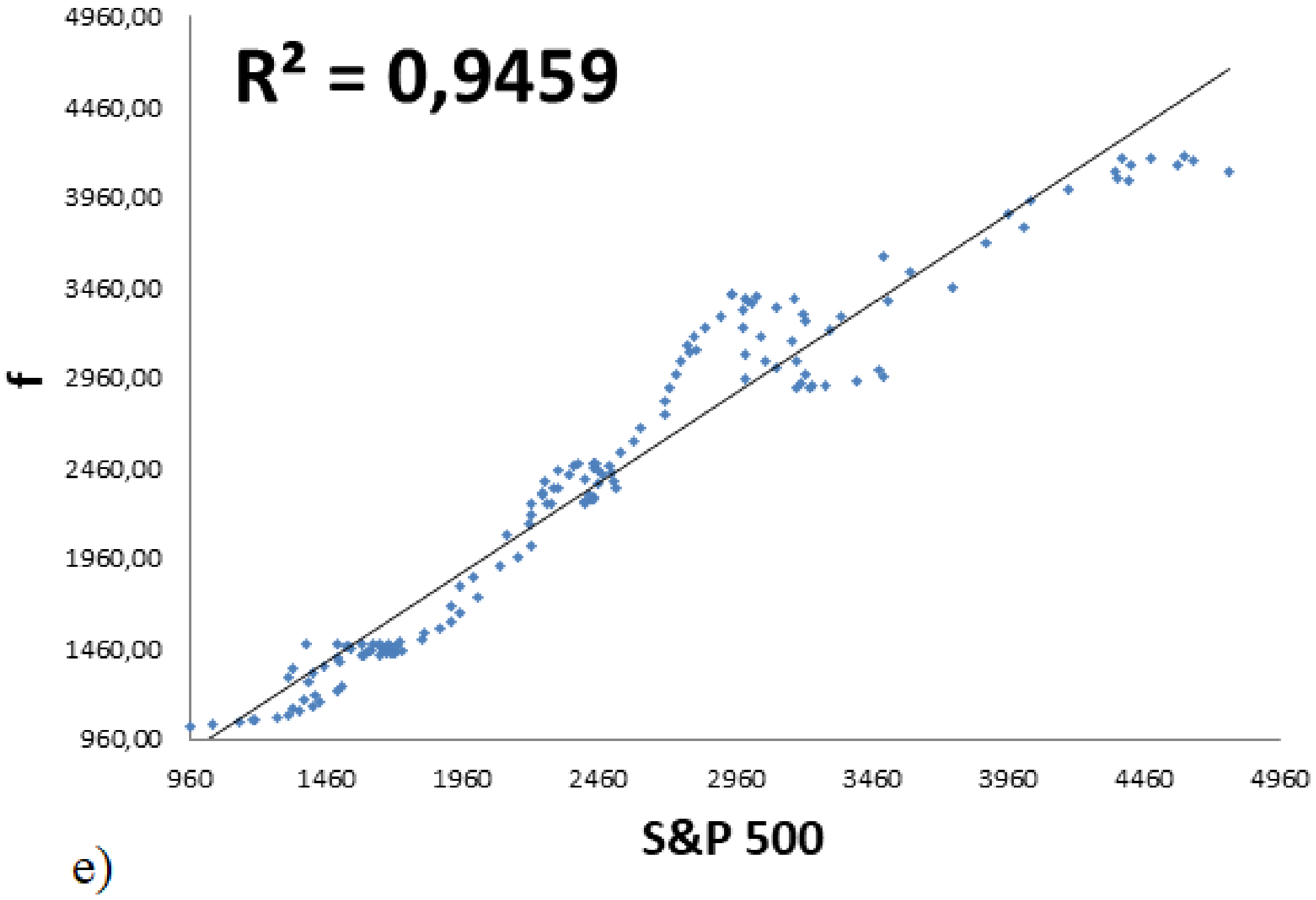

e)

**Figure 5:** Scatter plot of the a) Australian dollar (AUD/USD), b) Swiss franc (USD/CHF), c) Natural gas (NG), d) Corn, and e) S&P500 index values compared to model predicted values (*f*). Straight line – OLS fit.

**Table 2** OLS regression parameters (t-values are given in the parentheses)

| No | AUD/USD | USD/CHF | NG | Corn | S&P 500 |
|---|---|---|---|---|---|
| f | 1.01**<br>(155.43) | 1.00**<br>(-0.40) | 0.97**<br>(41.03) | 0.95**<br>(7.26) | 0.95**<br>(4.03) |
| Constant | -0,004**<br>(-4.18) | -0.002***<br>(195.03) | 0.09**<br>(1.36) | 52.39**<br>(64.03) | 190.56**<br>(51.89) |
| Observations | 690 | 1601 | 246 | 241 | 156 |
| Adjusted $R^2$ | 0.97 | 0.96 | 0.87 | 0.94 | 0.95 |

*p < 0.5; **p < 0.01; ***p > 0.5

It can be noted that the empirically observed data are linearly related to the values predicted by the model with a reasonable degree of accuracy. The ordinary least squares method allows the model to explain 88 – 96% of the variation. The OLS coefficients are close to unity for all listed market assets, indicating that the fitted model "performs well". The constants are insignificant from zero for all assets except corn and the S&P 500. However these constants can be considered small percentages because the scales for corn and S&P 500 are much broader than the scales for the rest of the assets.

To eliminate the problem of spurious correlation between the asset price and model predicted curve, one can perform a two-stage Engel - Granger co-integration test (Engel & Granger, 2015) by regressing the asset price using approximation function *f* and intercept, and then running the Augmented Dickey-Fuller test (ADF) for stationarity. Table 3 presents critical values of the Dickey–Fuller t-distribution. The test results are summarized in Table 4.

**Table 3** Critical values for Dickey–Fuller t-distribution

| | Without trend | | With trend | |
|---|---|---|---|---|
| Sample size | 1% | 5% | 1% | 5% |
| N = 25 | −3.75 | −3.00 | −4.38 | −3.60 |
| N = 50 | −3.58 | −2.93 | −4.15 | −3.50 |
| N = 100 | −3.51 | −2.89 | −4.04 | −3.45 |
| N = 250 | −3.46 | −2.88 | −3.99 | −3.43 |
| N = 500 | −3.44 | −2.87 | −3.98 | −3.42 |
| N = ∞ | −3.43 | −2.86 | −3.96 | −3.41 |

(Source: Fuller, 1976, p.373)

**Table 4** Unit root test results (ADF)

| Variables | Critical value 1% | Critical value 5% | Computed value |
|---|---|---|---|
| AUD/USD | -3.46 | -2.88 | -3.97 |
| USD/CHF | -3.43 | -2.86 | -4.23 |
| NG | -3.46 | −2.88 | -3.51 |
| Corn | -3.46 | −2.88 | -3.92 |
| S&P 500 | -3.46 | −2.88 | -3.20 |

Comparing the estimated and critical values, one can conclude that the null hypothesis of a unit root can be rejected at the 5% level for S&P 500 and at the 1% level for AUD/USD, USD/CHF, NG and Corn.

## VI. Discussion

The results show that the model can accurately describe certain patterns on price charts, such as trends, and can indicate whether a trend is likely to continue or terminate. The model can also be applied to predict not only future asset prices, but also economic crises. For example Standard and Poor's 500 stock market index includes the 500 largest companies whose shares are traded on US stock exchanges. This is one of the largest stock indices which largely reflects the state of the US economy (and the world economy). A model predicting the end of the third wave of a historical trend could indicate that we are on the cusp of the next global recession. It is also possible to test the behaviour of market asset prices during periods of price consolidation since Equation 7 also has periodic solutions in the form of cnoidal waves, but due to the paper volume limitations, it remains a topic of future research. In this paper a non-linear differential equation is considered as a basic technique for describing price dynamics. Differential equations were previously used by Caginalp & Balenovich to provide foundation to the technical analysis. The authors noted that "While economic and financial scholars often ignore or downplay the role of conventional technical analysis, financial experts are often quite eager to implement it without regard to the nature of the economic assumptions inherent in the methods” (Caginalp & Balenovich, 2003, p.7). They showed that some technical analysis patterns can be modelled using differential equations by assuming interactions between two or more groups of traders with different valuations and/or different motivational characteristics. However, the reasons behind the investor sentiment were not given.

The information approach allows us to establish another connection with technical analysis. This refers to Fibonacci retracement, which is used in technical trend analysis of financial market trends to predict price support and resistance levels. Fibonacci numbers are a sequence in which each subsequent number is the sum of the two previous numbers: 0, 1, 1, 2, 3, 5, 8, 13, 21 … One can construct the ratios of these numbers and mirror ratios. The most commonly used values are: 0.382; 0.500; 0.618; 1; 1.618; 2.618; 4.236. These ratios are often used in Fibonacci retracement to determine support and resistant levels and compare market price movements to each other (e.g. Colby, 2003). In an attempt to describe time evolution of the market, Miner (2002) used Fibonacci ratios to determine market time cycles. One common approach is to take past waves and use Fibonacci ratios to predict the relationship of past waves to future waves. For example using alternate price projection (APP) one can forecast the proportion of a past wave to the next wave, moving in the same direction. The most important comparison ratios for such swings that are considered to have the highest likelihood of supporting or breaking a trend are: 62%, 100%, 162%, 200%, 262%, and 424%. Another method involves the use of price expansion (Exp), which expands the price range of a wave. The same the most important ratios to use for price expansions are: 62%, 100%, 162%, 200%, 262%, and 424% (Miner, 2002). The reason which may make traders feel uncomfortable is that Fibonacci retracement is not grounded in any logical proof and there is no rational basis for this approach. The information approach may provide possible explanation for the origin of Fibonacci relationships. Table 5 lists the first seven Fibonacci ratios and the corresponding soliton amplitudes obtained from Equation 8. From Table 5 it appears that the Fibonacci ratios correspond relatively well to the soliton amplitudes with a maximum relative difference of 12%.

**Table 5**. First seven Fibonacci ratios and corresponding soliton amplitudes

| No | Fibonacci ratios | soliton amplitudes | difference (%) |
|---|---|---|---|
| 1 | 1 | 1 | 0 |
| 2 | 1.62 | - | - |
| 3 | 2.62 | 3 | 12.7 |
| 4 | 4.24 | 4 | 5.7 |
| 5 | 6.85 | 6 | 12.4 |
| 6 | 11.09 | 10 | 9.8 |
| 7 | 17.94 | 16 | 10.8 |

The connection between soliton amplitudes and Fibonacci ratios may shed light on the basis of Fibonacci number-based connection to Elliott waves.

The presented model can be used to predict trend reversals. At the very beginning, it is impossible to say for sure that a trend price movement will begin to unfold. But when the trend unfolds and wave structure is confirmed, one can expect to reach certain price levels at certain points in time. If the level is reached by the planned time point, a (temporary) trend reversal can be expected. However, the presented model should not be considered as a universal tool for forecasting financial time series in any periods of trend and price consolidation, since equation 7 was obtained under certain conditions that cannot always be met.

At the same time, the information approach contributes to the growing literature examining the role of social norms, moral attitudes, religions, and ideologies in the imperfect rationality of market participants. Recently, there has been a growing understanding that investors' thoughts and behavior are influenced by accepted cultural traits, which can be viewed as an evolutionary system. This forms the basis of a new paradigm for understanding financial markets – social finance (e.g. Akçay, Hirshleifer, 2020). There are significant similarities between social finance and information approaches. Table 6 summarizes some of the main characteristics of the two approaches.

**Table 6.** Social finance and Information approach

| Social finance | Information approach |
|---|---|
| Adopted cultural traits, including information signals, beliefs, strategies, and folk economic models are the drivers of investors' decisions | Sets of communication codes are the drivers of information processing and investors' decisions |
| Social transmission determine the evolution and mutation of financial traits | Sets of communication codes shape each other in the process of communication |
| Cultural traits are subject to different biases in judgments and decisions. | Different sets of communication codes provide different meanings to the same informational content |
| Cultural traits are not immediately measurable | Meaning can be measured on the base of the extension of Shannon's mathematical theory of |
| | communications |
| Social finance encompasses agent-based modeling of transmission biases | Information approach encompasses modelling the effect of meaning generation with help of non-linear evolutionary equation |
| Regards shifts in investors' sentiment as an endogenous outcome of microevolutionary cultural processes | Regards shifts in investors' sentiment as an endogenous outcome of interaction of communication codes sets, formed in the evolutionary process |
| Investors adopt and modify their financial traits | Agents adopt and modify their sets of communication codes |
| Considers a wider set of applications and range of time scales | Considers a wider set of applications and range of time scales |
| Cultural evolution operates at multiple time scales. The evolutionary dynamics of financial traits play out at multiple time scales | Communication code sets operate at multiple time scales. The evolutionary dynamics of meaning generation play out at multiple time scales |
| Is focused financial sphere | Can be applied a wider range of applications |

Like social finance, the information approach complements the study of how social interaction shapes the thinking and behavior of market participants. Both social finance and information approaches consider group characteristics of agents as the main cause of social transmission biases at different time scales. These features develop in the process of communication. There are also differences between the two approaches: 1) social finance is explicitly focused on how social interaction shapes the thinking and behavior of traders, the information approach goes beyond the field of finance and can be applied to the broader field of inter-social communications. (innovation research, rumors propagation, etc.); 2) cultural traits cannot be measured directly, whereas meaning can be measured using redundancy within the information approach; 3) social finance and information approach use different mathematical apparatus. The introduction of the method of nonlinear evolutionary equations allows us to adequately describe some observed price patterns and additionally provide a basis for technical analysis, including Fibonacci retracement.

## VI. Conclusion

This paper presents the first study of the dynamics of market price dynamics which examines the role of information processing and meaning generation in social systems. The study builds on the Loet Leydesdorff's seminal works on the dynamics of expectations and meaning in inter-human communications (Leydesdorff, 2008; Leydesdorff & Dubois, 2004; Leydesdorff, Dolfsma, Van der Panne, 2006; Leydesdorff & Franse, 2009; Leydesdorff & Ivanova, 2014; Leydesdorff, Petersen & Ivanova, 2017). It also incorporates the conceptual framework of the Triple Helix model of university-industry-government relations (Etzkowitz & Leydesdorff, 1995, 1998) and its mathematical formulation (Ivanova & Leydesdorff, 2014a, 2014b).

The main contribution of the paper is that it represents a step forward in the development of a theory of meaning that has potentially has many more arears of applicability than financial research. In relation to the study of stock and financial markets market assets' price movement is analysed from the viewpoint of information and meaning dynamics among market participants. This paper can also be considered an example of the application of the theory of meaning, which has long been considered an abstract discipline, to the specific field of financial markets. Another contribution is the introduction of a quantitative model to study the dynamics of market asset price movements and predict future price movements, based on a non-linear evolutionary equation.

Meaning in social communications is processed via specific sets of communication codes that span horizons of meaning, acting as selection and coordination mechanisms. It is based on expectations and caused by events against the arrow of time. Expectations can be measured as redundancy (i.e. additional options) using Shannon's entropy information theory.

Conceptually, there is much to be gained from combining and testing these ideas with new methods and approaches. Perhaps even more important is the idea that meaning can be quantified. This is a very interesting and promising research area that has great potential to expand our ability to understanding the dynamics of social systems and improve our ability to predict events that have not yet occurred. It can be useful when applied to various fields related to information exchange in social systems in general and to behavioral economics and financial markets in particular. The findings of the paper may also be useful for market practitioners in their daily activities.

The main provisions of the paper are as follows: the market can be considered as an ecosystem with bi- and trilateral relations between agents. In this respect mechanisms driving market evolution are similar to those of the TH model of innovations. Asset price dynamics can be analyzed from an information theory perspective, taking into account the relationships between information processing and meaning generation. Agents represent three groups of traders, which can be considered distributions spanning a network of relations through which information is communicated. There are other dynamics on the top of this network. Different agents use different communication codes, reflecting their preferences, to assign meaning to the information. Codes can be considered eigenvectors in a vector-space (von Foerster, 1960) and structure communications as selection environments. Communicated information is supplied with different meaning by different agents. Meaning is provided from the perspective of hindsight. Meanings cannot be communicated, as in case of information, but only shared. Providing information with meaning increases the number of options (redundancy). This mechanism can be considered probabilistically using Shannon's equations (Shannon, 1948). The generation of options (redundancy) is crucial for system change. The trade-off between the evolutionary generation of redundancy and the historical variation that provides uncertainty can be measured as negative and positive information, respectively. The dynamics of information, meaning, and redundancy can be evaluated empirically using the sign of mutual information as an indicator. When the dynamics of expectations, that generates redundancy, prevail over the historical construction that generates entropy, mutual redundancy is negatively signed because the relative uncertainty is reduced by increasing the redundancy. The balance between redundancy and entropy can be depicted in terms of two vectors (*P* and *Q*) which can also be understood in terms of the generation and reduction of uncertainty in communication resulting from the interactions between three (bi-lateral) communication channels. Eventually non-linear mechanisms in redundancy evolution may prevail leading to the predictable behavior in the evolution of market prices.  This does not mean that market price movements are completely predictable, but during certain periods a plausible estimate of price changes can be made.

The approach entertained in this paper considers the market an evolving social system and follows Luhmann's conjecture (Luhmann, 1982) that evolution theory, systems theory, and communication theory can be programmatically integrated from a sociological perspective.
The significance of paper findings is due not only to accurate description of some observed market phenomena but also to the connection with technical analysis and social finance. The noted connections of information approach with various arears of financial research and accepted practice of market practitioners give hope for the future unification of various approaches in market research. Future research includes applying the present approach to some other problems and datasets with different numbers of features.

The paper was prepared within the framework of the Basic Research Program at the National Research University Higher School of Economics.

**Acknowledgment**

The author is grateful to Loet Leydesdorff for his valuable comments on an earlier draft of this paper.

## Appendix A

Shannon informational entropy is defined as:

$$H = -\sum_{i=1}^{S} p_i log\, p_i \qquad \text{(A1)}$$

Dubois shows (2019) that taking into account temporal cyclic systems:

$$H = H(t) = -\sum_{i=1}^{S} p_i(t) log\, p_i(t) \qquad \text{(A2)}$$

with normalization conditions:

$$\frac{1}{T}\int_0^T \sum_{i=1}^{S} p_i(t)dt = 1, \qquad H_0 = \frac{1}{T}\int_0^T H(t)dt \qquad \text{(A3)}$$

in one can obtain harmonic oscillator equation for probabilities. Following Dubois one can define the reference state:

$$I_0 = -\sum_{i=1}^{S} p_{i,0} log\, p_{i,0} \qquad \text{(A4)}$$

and develop informational entropy $H$ in Taylor's series around the reference state:

$$H = I_0 - \sum_{i=1}^{S}[(logp_{i,0} + 1)(p_i - p_{i,0}) + \frac{(p_i - p_{i,0})^2}{2p_{i,0}} + \cdots O\left(\left(p_i - p_{i,0}\right)^3\right)] \qquad \text{(A5)}$$

substituting equation A4 into Equation A5 and neglecting the terms beyond the second degree we obtain:

$$H = -\sum_{i=1}^{S}[p_i logp_{i,0} + \left(p_i - p_{i,0}\right)] + D^* \qquad \text{(A6)}$$

where:

$$D^* = \sum_{i=1}^{S}\left(\frac{\left(p_i - p_{i,0}\right)^2}{2p_{i,0}}\right) \qquad \text{(A7)}$$

The condition for non-asymptotic stability of cyclic system is:

$$\frac{dD^*}{dt} = 0 \tag{A8}$$

In case $H(t) = H_1(t) + H_2(t) = -p_1 logp_1 - p_2 logp_2$ the solution to Equation A6 is:

$$\begin{cases} \frac{dp_1}{dt} = -\frac{\gamma}{p_{2,0}}(p_2 - p_{20}) \\ \frac{dp_2}{dt} = \frac{\gamma}{p_{10}}(p_1 - p_{10}) \end{cases} \tag{A9}$$

Differentiation of system A9 by time gives:

$$\begin{cases} \frac{d^2p_1}{dt^2} = \frac{-\gamma^2}{p_{10}p_{20}}(p_1 - p_{10}) \\ \frac{d^2p_2}{dt^2} = \frac{-\gamma^2}{p_{10}p_{20}}(p_2 - p_{20}) \end{cases} \tag{A10}$$

The function $D^*$ corresponds to the non-linear residue in (A6) which is a truncated version of (A5). For non-truncated equations (A5) we obtain:

$$\begin{cases} \frac{d^2p_1}{dt^2} = \frac{-\gamma^2}{p_{10}p_{20}}(p_1 - p_{10}) + C_1 \\ \frac{d^2p_2}{dt^2} = \frac{-\gamma^2}{p_{10}p_{20}}(p_2 - p_{20}) + C_2 \end{cases} \tag{A11}$$

where $C_j = O\left(\left(p_j - p_{j,0}\right)^3\right)$, j=1,2. When $p_1$ is less than $p_2$, to maintain the same order of magnitude we can eliminate the terms beyond the second power for the variable $p_1$ and the terms beyond the third degree for the variable $p_2$. Similarly manner this leads to the function $D^{**}$ defined by analogy with $D^*$:

$$D^{**} = \frac{(p_1-p_{10})^2}{2p_{10}} + \frac{(p_2-p_{20})^2}{2p_{20}} - \frac{(p_2-p_{20})^3}{6{p_{20}}^2} \tag{A12}$$

Differentiating $D^{**}$ by time and equating to zero $\frac{dD^{**}}{dt} = 0$ we obtain the system:

$$\begin{cases} \frac{dp_1}{dt} = \frac{\gamma}{p_{20}}(p_2 - p_{20}) - \frac{\gamma}{2{p_{20}}^2}\left(p_2 - p_{2,0}\right)^2 \\ \qquad \frac{dp_2}{dt} = -\frac{\gamma}{p_{10}}(p_1 - p_{10}) \end{cases} \tag{A13}$$

which gives the equation for a non-harmonic oscillator:

$$\frac{d^2p_2}{dt^2} = -\frac{\gamma^2}{p_{10}p_{20}}(p_2 - p_{20}) + \frac{\gamma^2}{{p_{20}}^2 p_{1,0}}(p_2 - p_{20})^2 + C_1 \qquad \text{(A14)}$$

If $H(t) = H_1(t) - H_2(t)$ (that is when self-organization takes place, as described by equation 5) the non-harmonic oscillator equation is:

$$\frac{d^2p_2}{dt^2} = \frac{\gamma^2}{p_{10}p_{20}}(p_2 - p_{20}) - \frac{\gamma^2}{{p_{20}}^2 p_{1,0}}(p_2 - p_{20})^2 + C_2 \qquad \text{(A15)}$$

## Appendix B

Redundancy (Equation 5) can be viewed as the result of a balance between two dynamics - historical organization and evolutionary self-organization (Leydesdorff, 2010). In other words it is a balance between recursion on a previous state on the historical axis and the meaning provided to the events from the perspective of hindsight (Dubois, 1998). Redundancy dynamics govern corresponding probability dynamics from a recursive and incursive perspectives. Resulting dynamics is provided by two tendencies - historical organization and evolutionary self-organization. Assuming that probabilities oscillate in a non-harmonic mode (Equations A14 and A15) one can write:

$$\frac{d^2p}{dt^2} = \frac{\gamma^2}{p_{10}p_{20}}(p^- - p) - \frac{\gamma^2}{{p_{20}}^2 p_{10}}(p^- - p)^2 - \frac{\gamma^2}{p_{10}p_{20}}(p^+ - p) + \frac{\gamma^2}{{p_{20}}^2 p_{10}}(p^+ - p)^2 + C_1 + C_2 \quad \text{(B1)}$$

here $p_j{}^-$ and $p_j{}^+$ are defined with respect to past and future states.

Developing $p^+$ and $p^-$ in the Taylor's series in state space:[10]

$$\begin{aligned} p^+ &= p + p'h + \frac{1}{2}p''h^2 + \frac{1}{6}p'''h^3 + \frac{1}{24}p''''h^4 + \cdots \\ p^- &= p - p'h + \frac{1}{2}p''h^2 - \frac{1}{6}p'''h^3 + \frac{1}{24}p''''h^4 + \cdots \end{aligned} \quad \text{(B2)}$$

and keeping the terms up to the $h^4$ order of magnitude one obtains (Fermi, Pasta, Ulam, 1955):

$$\frac{1}{k}p_{tt} = p''h^2 + 2\alpha p'p''h^3 + \frac{1}{12}p''''h^4 + O(h^5) \quad \text{(B3)}$$

here: $k = \frac{\gamma^2}{p_{10}p_{20}}$; $\alpha = \frac{1}{p_{10}}$; $C = \frac{1}{k}(C_1 + C_2)$. Setting further: $w = \sqrt{k}$; $t' = wt$; $y = {}^{x}/_{h}$; $\varepsilon = 2\alpha$ one can rewrite equation B3 in the form:

$$-p_{\tau\tau} + p_{yy} + \varepsilon p_y p_{yy} + \frac{1}{12}p_{yyyy} + C = 0 \quad \text{(B4)}$$

---

[10] state space is presented by x axis

Passing to a moving coordinate system: $X = y - t'$, rescaling the time variable $T = \frac{\varepsilon}{2}\tau$, and retaining terms to first order in $\varepsilon$ equation B3 takes the form:

$$\varepsilon\Sigma_{XT} + \varepsilon\Sigma_X\Sigma_{XX} + \frac{1}{12}\Sigma_{XXXX} + C_1 = 0 \qquad \text{(B5)}$$

Here $p_j = \Sigma(\mathrm{X}, \mathrm{T})$. Defining further: $P = \Sigma_X$ and $\delta = \frac{1}{12\varepsilon}$ we obtain a non-linear evolutionary equation:

$$P_T + PP_X + \delta P_{XXX} + C_1 = 0 \qquad \text{(B6)}$$

which is Korteweg de Vries (KdV) equation (Gibbon, 1985*)*:

$$u_T + uu_X + \delta u_{XXX} = 0 \qquad \text{(B7)}$$

with additional term $C_1$. By substitution: $u \to 6U;\ T \to \sqrt{\delta}\tau;\ \ X \to \sqrt{\delta}\chi$ Equation B7 is reduced to the form:

$$U_\tau + 6UU_\chi + U_{\chi\chi\chi} = 0 \qquad \text{(B8)}$$

equation. B8 has soliton solutions:

$$U(\chi, \tau) = \frac{\kappa^2}{2} sech^2\left[\frac{k}{2}(\chi - \kappa^2\tau)\right] \qquad \text{(B9)}$$

soliton solution to equation B6 is:

$$P(X, T) = \frac{\kappa^2}{12} sech^2\left[\frac{k}{2\sqrt{\delta}}\left(X - \kappa^2 T + \frac{C_1}{2\sqrt{\delta}}T^2\right)\right] - C_1 T \qquad \text{(B10)}$$

It follows that for a sequence of solitons with amplitudes: $A_i$, $A_j$ and $A_k$ and times $T_i$, $T_j$, $T_k$ the following condition is satisfied: $\frac{A_i - A_j}{T_i - T_j} = \frac{A_j - A_k}{T_j - T_k}$

**Appendix C**

Assuming the existence of a periodic solution (Gibbon, 1985):

$$U(x,t) = f(x - vt) \tag{C1}$$

substituting expression C1 into equation B7 gives:

$$-vf' + ff' + f''' = \frac{1}{2}a \tag{C2}$$

here $a$ is the integration constant. Multiplying the equation C2 by $2f'$ and integrating it we obtain an equation that has periodic solutions in the form of elliptic functions:

$$f'^2 = -\frac{1}{3}f^3 + vf^2 + af + b \tag{C3}$$

Then the corresponding solution to equation B6 will take the form:

$$P(x,t) == f\left(x - vt + {}^{1}/_{2}\, Ct^2\right) - Ct \tag{C4}$$